\documentclass[journal=jacsat,manuscript=article]{achemso}
\usepackage{chemformula} % Formula subscripts using \ch{}
\usepackage[T1]{fontenc} % Use modern font encodings

\author{Luca Assogna}
\author{Giovanna Salvitti}
\author{Matteo  Silvestri}
\author{Federico Perrella} 
\affiliation{Dipartimento di Scienze Fisiche e Chimiche, Università degli Studi dell’Aquila,Via Vetoio, 67100 L’Aquila, Italy}
\author{Riccardo Mincigrucci} 
\author{Cristian Soncini} 
\affiliation{Elettra-Sincrotrone Trieste S.C.p.A., SS 14-km 163.5, 34149 Basovizza, Trieste, Italy}
\author{Elena Incerto}
\author{Armando Carlone}  
\affiliation{Dipartimento di Scienze Fisiche e Chimiche, Università degli Studi dell’Aquila,Via Vetoio, 67100 L’Aquila, Italy}
\author{Majed Chergui} 
\affiliation{Elettra-Sincrotrone Trieste S.C.p.A., SS 14-km 163.5, 34149 Basovizza, Trieste, Italy}
\alsoaffiliation{Lausanne Centre for Ultrafast Science (LACUS), ISIC-FSB, Ecole Poleytechnique Fédérale de Lausanne, 1015 Lausanne, Switzerland}
\author{Claudio Masciovecchio}
\affiliation{Elettra-Sincrotrone Trieste S.C.p.A., SS 14-km 163.5, 34149 Basovizza, Trieste, Italy}
 \author{Paola Benassi}
 \author{Andrea Marini}
 \affiliation{Dipartimento di Scienze Fisiche e Chimiche, Università degli Studi dell’Aquila,Via Vetoio, 67100 L’Aquila, Italy}
 \alsoaffiliation{CNR-SPIN, c/o Dip.to di Scienze Fisiche e Chimiche - Via Vetoio, 67010, L'Aquila, Italy}
  \author{Davide Tedeschi} 
  \affiliation{Dipartimento di Scienze Fisiche e Chimiche, Università degli Studi dell’Aquila,Via Vetoio, 67100 L’Aquila, Italy}
  \email{davide.tedeschi@univaq.it}
  \author{Carino Ferrante} 
  \affiliation{CNR-SPIN, c/o Dip.to di Scienze Fisiche e Chimiche - Via Vetoio, 67010, L'Aquila, Italy}
\email{carino.ferrante@spin.cnr.it}

\title[]
  {Femtosecond self-diffraction as a measure of the nonlinear response spectrum}
\abbreviations{IR,NMR,UV}
\begin{document}
\begin{abstract}
	Self diffraction is a four-wave mixing process proportional to the square modulus of third-order nonlinearity susceptibility $\chi^{(3)}$, which is related to the material's electronic and thermal properties.
	In this study, we investigate the wavelength dependence of the self-diffracted signal generated by a femtosecond pulsed laser in a dye solution to directly evaluate the electronic third-order nonlinear susceptibility spectrum. By accounting for absorption effects and phase matching conditions, we determine the $\vert\chi^{(3)}\vert$ for different concentrations. Experimental results complemented with theoretical predictions, show that in the low absorption and thin sample limits, the signal reproduce the $\vert\chi^{(3)}\vert$ spectral profile.
	These findings demonstrate the feasibility of measuring nonlinear susceptibility spectra arising solely from the bound-electronic response across a wide spectral range and for various compounds.
\end{abstract}

In recent years, nonlinear optics has opened new pathways for manipulating light–matter interactions at the nanoscale and enabling a wide range of applications. Some of them include all-optical signal processing as well as the generation of entangled photon pairs for  quantum communication protocols ~\cite{solis2021dependence,zhao2021nonlinear,gu2016molecular,trovatello2025quasi}. For this purpose, the identification of the appropriate material with a substantial effective nonlinear susceptibility within a specified spectral range
is essential for the efficient implementation of all-optical modulation. 
A number of techniques enable the experimental determination of the third order susceptibility ($\chi^{(3)}$) of a given medium, such as \emph{Z}-scan ~\cite{liaros2017characterization,de2016techniques}, beam deflection ~\cite{reichert2014temporal,zhao2018temporal}, and four-wave mixing (FWM) techniques ~\cite{maidur2021ultrafast,kumar2009ultrafast}. In the field of material science, \emph{Z}-scan has been extensively adopted for the characterization of diverse compounds~\cite{lee2021investigation,arunsankar2025solvent}.
This is primarily attributed to its relative simplicity in comparison to alternative methods. 
However, such an experimental approach necessitates the scanning of the sample's position along the beam's focus, namely \emph{Z}-values, for each wavelength excitation, hindering spectral investigation and providing a limitation for
fragile samples, as it requires multiple laser shots to obtain the \emph{Z}-scan profile.
Conversely, it should be noted that beam deflection experiments are limited to materials where the absorption is sufficiently small so that electromagnetic radiation throughout the sample is constant, and a full description of the beam shape propagating through optics is mandatory~\cite{ferdinandus2017analysis,Ferdinandus:13}. 
Finally, the wavelength scan of a FWM signal, depending on the specific implemented process, is subject to intensity modulation, based on the 
phase-matching conditions (PMCs), and beam direction modulation, due to the sum of photon momenta.
These effects prevent an easy spectral detection of $\chi^{(3)}(\lambda)$. Among the FWM schemes, one is transient grating (TG), in which two degenerate pump beams induce a spatial transient grating by interference effects, which a probe pulse can monitor being deflected in a specific direction~\cite{Hoffman:86}.

This study illustrates a  methodology for measuring  $|\chi^{(3)}(\lambda)|$, through a self-diffraction (SD) scheme. 
SD can be described as the degenerate case of TG, whereby one pump beam probes the action of itself through the nonlinear medium~\cite{1989_newmethodpulsetimechar,Trebinopulsed_cw,wiese1989nonlinear}.
Different from TG experiments, the SD technique is not capable of studying the dynamics of the transient excitation process due to the absence of a delayed probe beam. Moreover, the wavelength of the probe beam being equivalent to that of the pump does not allow to completely fulfill the PMCs. 
In ref.~\cite{wiese1989nonlinear} a SD scheme was employed to measure the nonlinear susceptibility of Cresyl violet ($\rm C_{19}H_{18}ClN_{3}O$) with light pulses of 20 ps. 
However, the effect of the pulses' impinging angle was not taken into account. Moreover, the utilization of ps pulses was also unable to fully disentangle significant non-instantaneous mechanisms due to nuclear motions on the nonlinear medium response ~\cite{reichert2014temporal,zhao2018temporal}. 
This may result in an overestimation of the electronic $|\chi^{(3)}(\lambda)|$ value, as observed by means of \emph{Z}-scan measurements conducted on the same Methyl Blue molecule (CAS no. 28983-56-4, BLDpharm) under investigation in the present study ~\cite{ling2023z,parvin2015nonlinear}. Therein, the transition from the femtosecond pulses to the continuous wave regime induces an approximate $7$ orders of magnitude difference in the estimation of the $|\chi^{(3)}(\lambda)|$ value.
Therefore, it is imperative to employ a femtosecond pulsed laser to suppress any non-electronic contributions to the nonlinear response.

\begin{figure}[t!]
	\centering
	\includegraphics[width=  0.55\linewidth]{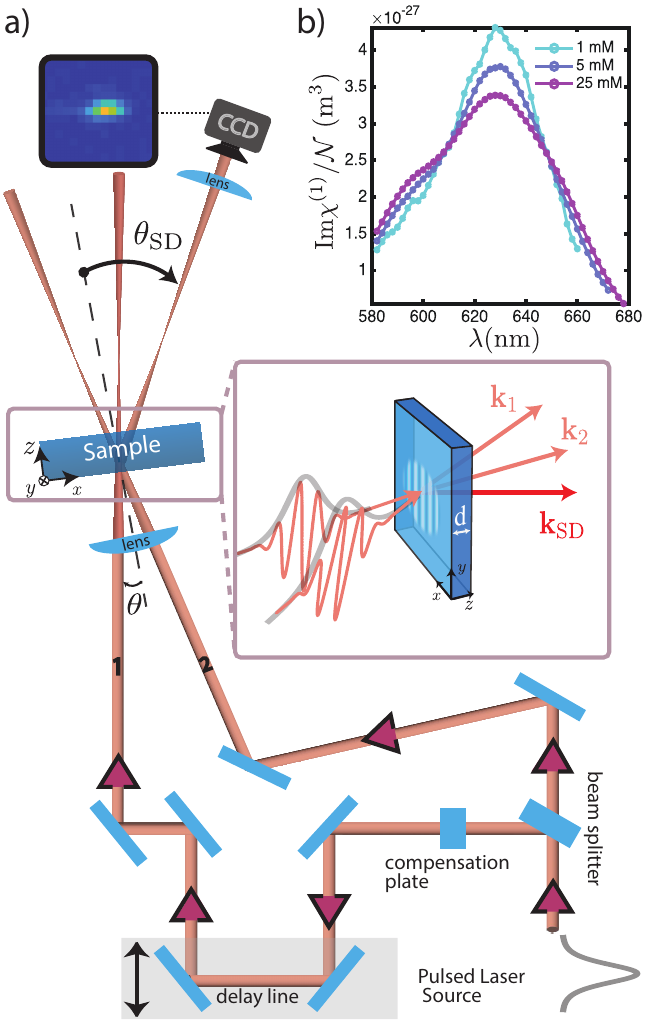} % aggiornata al 3 ottobre  
	\captionsetup{justification=justified, singlelinecheck=false}
	\caption{ 
		{\bf a)} SD experimental setup scheme. Starting from the bottom, the fs pulse emitted by an OPA enters the optical path. A 50:50 plate beam splitter divides the pump into two equivalent beams, $1$ and $2$. Beams $1$ and $2$ are focused by a lens with focal length of $10\ {\rm cm}$ before reaching a $10\ \mu \rm m$ thick sample with angle of incidence $-\theta$ and $\theta$, respectively. Time-overlap is tuned by a delay line in beam $1$ path. The temporal chromatic dispersion  on the two arms are balanced by a compensation plate.
		After the sample, a second lens with a focal length of $20\ {\rm cm}$ collimates the SD signal towards a CCD, whose typical example is shown in the colormap.
		The inset provides a visual representation of the SD process in our reference frame. {\bf b)} Measured imaginary part of the linear susceptibility normalized by the molecular number density ${\rm Im}\chi^{(1)}(\lambda)/\mathcal{N}$ as a function of the wavelength for three molar concentrations (C) of Methyl Blue dye solutions dispersed in water.}
	\label{fig1}
\end{figure}

The considered optical configuration for the SD experimental setup is illustrated in Figure \ref{fig1}a. 
The pulsed laser source (Pharos, PH2-20W-SP, Light Conversion) has a central wavelength of $1030 \pm 10 \rm nm$ and a pulse duration of approximately $190 \rm fs$.
Pulses with repetition rate of $10\  {\rm kHz}$ and energy of $180 \ \mu \rm J$ per pulse are employed to feed an optical parametric amplifier (OPA), to achieve wavelength-tunable pulses with $\tau_{\rm FWHM} \simeq 50\ {\rm fs} $. The emitted wavelength from the OPA (ORPHEUS-F, Light Conversion) is controlled by a sequence of motors, enabling an automated wavelength scan. Subsequent to the OPA a 50:50 fused silica beam splitter of thickness $l =$ 5 mm is positioned at an angle of $ 45^\circ$ with respect to the propagation direction. Considering the requirement of temporal overlap between the two beams, a delay line was utilized to obtain time overlap on the sample.
In order to achieve a time overlap condition independent on wavelength, 
the temporal dispersion  arising from the propagation through the beam splitter in arm $2$ is compensated by a fused silica compensation plate, inserted in arm $1$.
Subsequently, both beams are directed through a common spherical lens ($f=10\ {\rm cm}$) and impinge on the sample at the same spatial position, with incident at angles of $\pm\theta$ with respect to the sample's surface normal.
The sample is $10\ \mu \rm m$-thick Methyl Blue dispersed in water, contained in a quartz cuvette. The signal at $\theta_{SD}\simeq 3 \theta$ is then collimated by a lens with focal length of $20\ {\rm cm}$ and focused  towards a CCD camera (Basler acA780-75gm). 
		Within the framework of the proposed methodology, the SD signal maintains a consistent direction throughout a wavelength scan, thereby enabling the study of the SD signal as a function of wavelength without the necessity for adjustment of the setup geometry. Consequently, the measurement's typical duration is approximately a few minutes per spectrum. 
		
		\begin{figure}
			\centering
			\includegraphics[width=  1.03\linewidth]{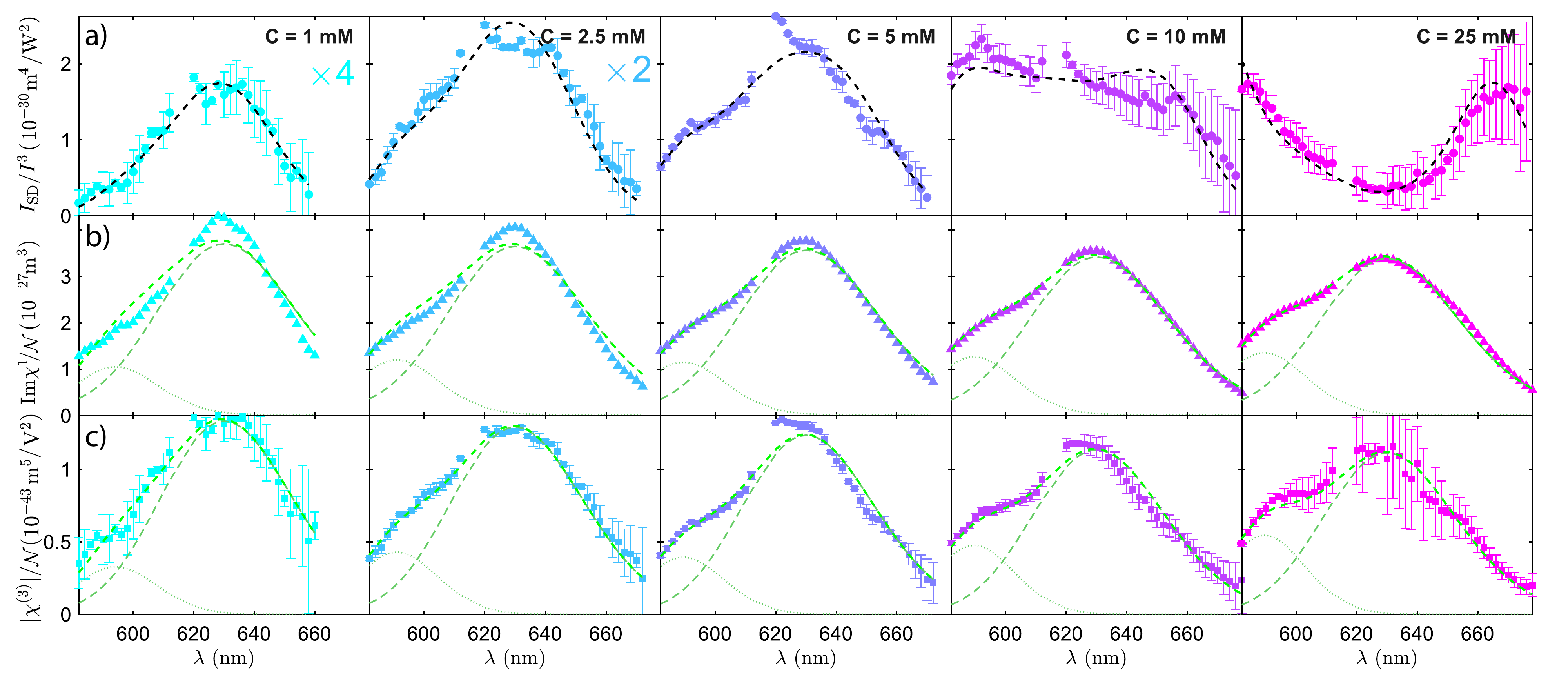}
			
			\captionsetup{justification=justified, singlelinecheck=false}
			
			\caption{{\textbf a)} Colored dots represent SD signals normalized by the peak intensity of the beams, acquired at steps of $2\ {\rm nm}$ for different C. Peak-to-valley dynamics are observed as C increases. The black lines are a fit to the data. {\textbf b)} Colored triangles represent $ {\rm Im}\chi^{(1)}(\lambda)$ spectra acquired linear regime, rescaled by number density $\mathcal{N}$. Green lines are fit to the data, while the green dotted and dashed lines represent the two Lorentzian components of ${\rm Im}\chi^{(1)}(\lambda)$ convoluted with the Gaussian contribution reported in \ref{fig1}b). \textbf{c)} Colored squares show the $\vert\chi^{(3)}\vert$ spectra, which were retrieved by rescaling the SD intensities using \eqref{eq:main}. The green lines show the sum of the two Gaussian-convoluted Lorentzian components of $\vert\chi^{(3)}\vert$, which are shown as the green dotted and dashed lines.}
			
			\label{main}
		\end{figure}
		The optimum position for the sample with respect to the focal point is a trade-off between two competing effects. The amount of nonlinear modulation increases with the peak intensity, see S.I., reaching the maximum at the  minimum spot size.
		Conversely, in the event of spot sizes being too small, a low number of interference fringes, due to the refractive index modulation, is sensed by the third probing interaction. Consequently, this generates a low-intensity SD signal, due to the low effective grating yield.  For this reason, the sample has been positioned slightly in front of the lens focal point ($1-2$ cm). A superior compromise could be attained combining cylindrical lenses.

		In this experiment, the measurements are conducted within a range of molar concentrations (C) of Methyl Blue in water ranging from $1\ {\rm mM}$ to $25 \ \rm {\rm mM}$. 
		In Figure \ref{fig1}b, the imaginary parts of the linear susceptibility spectra are depicted for varying  C. The transmission of diverse solutions was measured, and the imaginary part of the linear susceptibility was modeled by fitting the retrieved absorbance with two Gaussian components accounting for the absorption's inhomogeneous broadening. As C increases, the main component at $\sim 630$~nm becomes less intense with respect to the side component at $\sim 595$~nm. It can be deduced that this occurrence most likely arises from to the formation of aggregates within the solution, as observed in analogous dye molecules  \cite{aggreg1, aggreg2}. Such aggregates are the responsible of the side contribution.
		Consequently, in our model we introduce two effective transition dipole moments, to take into account the relative concentration between aggregates  and dispersed single molecules, and  it is reflected in the nonlinear response of the solution, see S.I.. 
		
		Figure \ref{main}a shows the intensity of the SD signal as colored dots taken at steps of $2 \rm \ {\rm nm} $, normalized by the peak intensity of the two beams.  The SD intensity is evaluated considering the counts of the signal in the CCD image (see insert in Figure \ref{fig1}a acquired with 1ms integration time) and rescaling it by the spectral profile of detector sensitivity, transmission of lenses, and Fresnel transmission coefficient induced by the sample interfaces. Further details on the procedure are reported in S.I.. The sample thickness $d=10$ $\mu {\rm m}$ and the beams' angles of incidence with respect to the sample's normal surface are $\theta_1=\theta$ and $\theta_2=-\theta$, with $\theta=3^\circ$.The C range extends from $1\ {\rm mM}$  (left most column) up to $25\ {\rm mM}$ (right most column). At low $C$ the signal exhibits a peak-like shape in correspondence with the absorption peak (at $\sim630$ \ {\rm nm}), while at higher $C$ the spectral profile assumes a valley-like shape. This observation can be rationalized by two competing  effects: the nonlinearity, which is responsible for the peak-like shape, and the linear absorption, responsible for the suppression of SD signal at $\sim630$ \ {\rm nm} and the consequent valley-like behavior. By increasing the concentration of the dye molecule, the absorption becomes increasingly detrimental.
		
		To reproduce our experimental data and extract accordingly the $|\chi^{(3)}(\lambda)|$, we perturbatively derive the SD signal generated via Kerr nonlinearity, see S.I., obtaining: 
		
		\begin{equation}
			I_{\rm SD}(\lambda) = \frac{9 \left| \chi^{(3)}(\lambda)  \right|^{2}}{4 \varepsilon_0^2 c^2} I_1^2(\lambda) I_2(\lambda) \left| \mathcal{F}(k_0,d,\varepsilon,\theta_1,\theta_2) \right|^2,
			\label{eq:main}
		\end{equation}
		where $\varepsilon_0$ is the vacuum permittivity, $c$ is the speed of light, $ I_1(\lambda)$ and $ I_2(\lambda)$ are the wavelength-dependent pump intensities, and $k_0 = 2\pi/\lambda$ is the vacuum wavenumber. In SD the role of PMCs and absorption, which is related to the linear light-matter response, are taken into account by
		\begin{equation}
			\mathcal{F}(k_0,d,\varepsilon,\theta_1,\theta_2) =k_0^2 \frac{ \left( e^{i\Psi d} - e^{i\phi d} \right)}{\Psi^2 - \phi^2},
			\label{eq:F_psi_phi}
		\end{equation}
		where \(\varepsilon(\lambda)  = (n + i\kappa)^2\) is the complex dielectric constant of the medium. The coefficients \(\phi\) and \(\Psi\) are given by
		\begin{align}
			\phi &= k_0 \left( 2\sqrt{\varepsilon - \sin^2\theta_1} - \sqrt{\varepsilon^* - \sin^2\theta_2} \right)=2 k_{1}^z - k_{2}^{z*},\
			\label{eq:phi}
		\end{align}
		\begin{align}
			\Psi &= k_0 \sqrt{ \varepsilon - (2\sin\theta_1 - \sin\theta_2)^2  }=k_{\rm SD}^z,
			\label{eq:psi}
		\end{align}
		where ${\bf k}_{1}$, ${\bf k}_{2}$, and ${\bf k}_{\rm SD}$ are the pump $1$, $2$, and self-diffracted wave vectors inside the sample, respectively.
		For $d\rightarrow0$ and negligible absorption, 
		$\mathcal{F} \propto k_0$. Differently, increasing $d$ the lack of PMC generates a strong suppression of the SD signal, together with a strong dependencies on impinging angles and wavelengths. As reported in the S.I., this model is crucial for the correct choice of the experimental condition in the employed setup
		correspond to a case with negligible dependence on PMC within the considered spectral range.
		Moreover, the imaginary part of dielectric constant - related to absorption and accounted for by  $\left|\mathcal{F}(k_0,d,\varepsilon,\theta_1,\theta_2)\right|^2$ - induces a suppression of the signal that follows an exponential proportionality with C.
		
		At this point, two possible approaches can be taken to extract the nonlinear susceptibility spectrum from SD signal. 
		By applying \eqref{eq:main}, the values of $\lvert\chi^{(3)}\rvert$ can be directly extracted from the SD measurements, as shown in Figure \ref{main}c by the colored squares. The retrieved spectra are obtained by properly rescaling the SD intensities by $\lvert\mathcal{F}\rvert^2$. The latter can be constructed from the measured dielectric permittivity of the sample $\varepsilon(\lambda)$, whose imaginary part is reported as colored triangles in Figure \ref{main}b, as detailed in the S.I..
		In the lower concentration regime, $\mathcal{F}$ is observed to be a slowly varying function, and the $\vert\chi^{(3)}\vert$ shape closely reflects the SD signal. Conversely, increasing C results in the development of a dip in $\mathcal{F}$, closely mirroring the shape of the SD signal. In this case, the proper determination of $\vert\chi^{(3)}\vert$ is achieved by taking into account  $\varepsilon(\lambda)$, given the significant correlation between $\mathcal{F}$ and $\varepsilon(\lambda)$.
		An alternative approach to retrieve  $\vert\chi^{(3)}\vert$ implies a complete unbiased fitting of the SD signal, where the molecular effective transition electric dipole  is treated as the fitting parameter.
		Specifically, each transition allows to calculate $\varepsilon(\lambda)$  and $\vert\chi^{(3)}\vert$ and, consequently, the SD signal can be calculated by \eqref{eq:main}. Then, by properly fitting the SD signal, a self consistent derivation of optical response of the system can be obtained.
		In agreement with the results reported in Figure \ref{fig1}b, we consider two electric dipoles to represent the optical response of the dye compound. By properly tuning the effective dipoles moments, we calculate the green dashed lines shown in Figure \ref{main}b, whose sum is in excellent agreement with the experimental results. As described in the S.I., the calculation of the optical parameters adopted to reproduce the data involves two Lorentzian contributions for ${\rm Im} \chi^{(1)} $ and $\lvert\chi^{(3)}\rvert$. Both are convolved with the Gaussian profile associated to each transition, showed as dotted line in Figure \ref{fig1}b, to account for inhomogeneous broadening (see S.I.). Each contribution is shown as a green line in Figure \ref{main}b–c.
		The excellent agreement between the $\vert\chi^{(3)}\vert$ spectra obtained using the two different methods  testifies the reliability of the procedure. However, the fitting procedure provides greater insight into the dynamics observed as a function of C in relation to the rescaling procedure.
		The validity of the fitting procedure is also confirmed by the almost linear proportionality with $\mathcal{N}$ of both 
		$\lvert\chi^{(3)}\rvert$ and $ {\rm Im} \chi^{(1)} $ components. Moreover, the evolution of the two Gaussian contributions with \( C \), reported in Figure~\ref{fig1}b and Figure~\ref{main}b-c, is well reproduced by the model. It is worth notice that in the limit of low C, or for a weakly absorbing medium, the determination of ${\rm Im}\chi^{(1)}$ becomes unnecessary, since the features of the $\lvert\chi^{(3)}\rvert$ spectrum closely follow those of the SD signal.
		\begin{figure}[t]
			\centering
			\includegraphics[width=  0.65\linewidth]{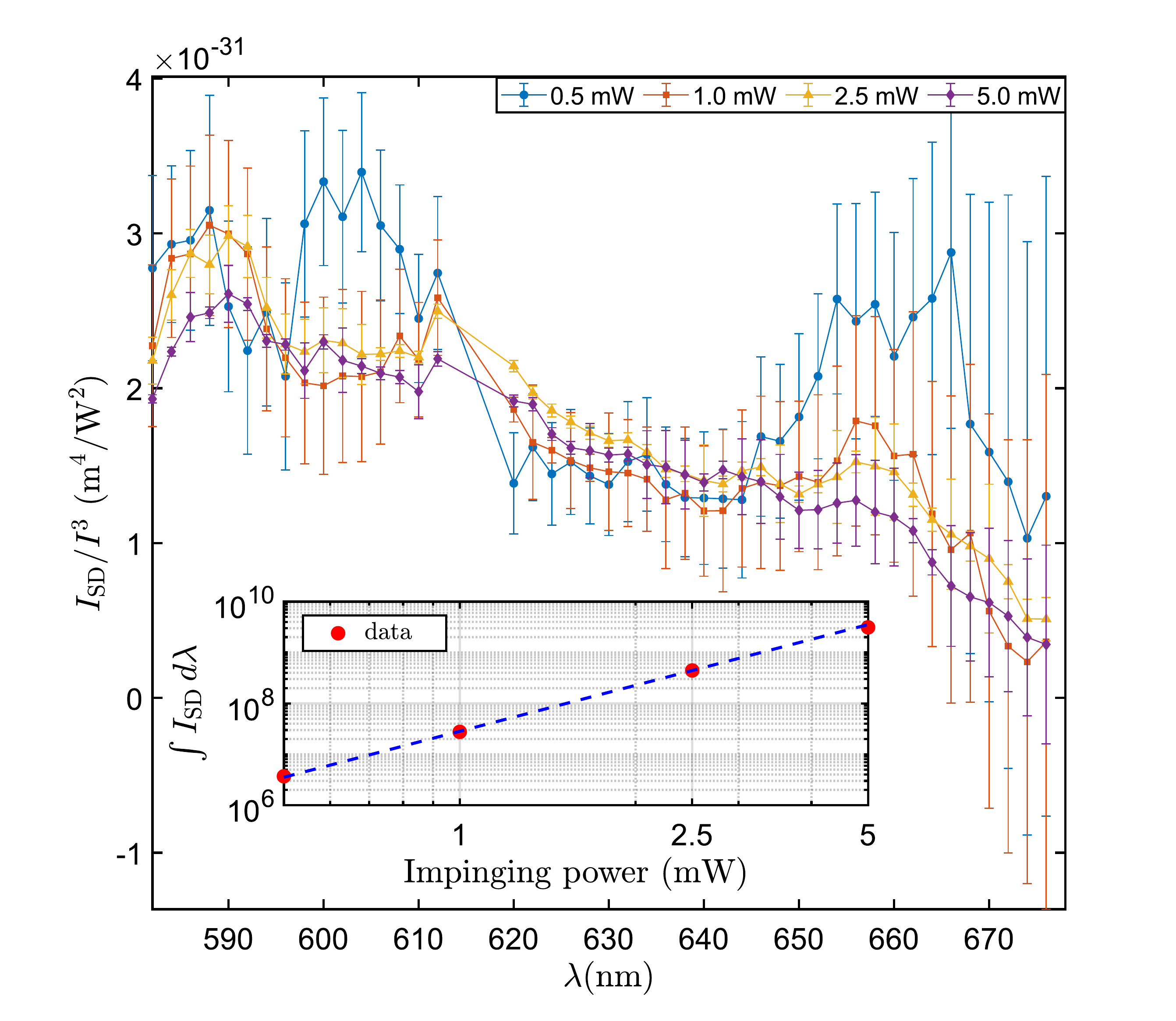}
			\captionsetup{justification=justified, singlelinecheck=false}
			
			\caption{SD signal as a function of the wavelength normalized by the peak intensity measured for different impinging power, measured before the beam splitter in Figure \ref{fig1}a at fixed wavelength $\lambda = 640\rm \ {\rm nm}$. The error associated with the 0.5 mW-power measurement is reduced by a factor 3 for sake of clarity. Insert: integrated SD signal as a function of the impinging power. The blue line is a third-power reference curve.}
			\label{fig:fig3}
		\end{figure}
		
		Finally, Fig \ref{fig:fig3} shows the SD signal, which has been normalized by the peak intensity of the two beams, as acquired at $C= 10\ {\rm mM}$ for several peak intensities. Considering the signal proportional to $ I^{3} $, the error bars decrease in conjunction with an increase in beam's power. As evidenced by the experimental results, the spectral shape of the normalized SD signal does not depend on the beams power. Moreover, the integrated SD signal is anticipated to adhere to a third-order power law, as expected by an homodyne third order effect. This dependence is shown in the insert in Figure \ref{fig:fig3}, where the integral over the wavelength of the SD signal  is reported as a function of the impinging power, thereby confirming the third-order nonlinear nature of SD. 
		
		In summary, the present study reports the measurement of the electronic third-order nonlinear susceptibility spectrum using a SD experimental setup. Furthermore, a model was developed to successfully reproduce the data and the corresponding PMC modulation, even when linear absorption has a strong detrimental effect on the nonlinear process. Such a model is fundamental for the choice of correct sample thickness, incidence angle, and concentration, able to suppress linear contributions and modulation induced by phase mismatch, providing a touchstone for the next design of SD experiment. The results demonstrate the capability of measuring the wavelength dependence of $|\chi^{(3)}(\lambda)|$, thus opening up new avenues for research that will facilitate a more comprehensive understanding of the role of resonances in nonlinear optical interactions.

\begin{acknowledgement}
This work has been partially funded by the European
Union-NextGenerationEU under the Italian Ministry
of University and Research (MUR) National Innovation
Ecosystem Grant No. ECS00000041—VITALITY—CUP
E13C22001060006.  
M. C. acknowledges support from the ERC Advanced Grant CHIRAX (n° 101095012), of the European Union's Horizon 2020 research.

\end{acknowledgement}

\begin{suppinfo}

The following file is available free of charge.
\begin{itemize}
  \item Analytical derivation of the SD intensity; Absorption effects in Methyl blue; From the Self diffracted signal to the $|\chi^{(3)}(\lambda)|$; Experimental Methods (PDF)
\end{itemize}

\end{suppinfo}

\providecommand{\latin}[1]{#1}
\makeatletter
\providecommand{\doi}
  {\begingroup\let\do\@makeother\dospecials
  \catcode`\{=1 \catcode`\}=2 \doi@aux}
\providecommand{\doi@aux}[1]{\endgroup\texttt{#1}}
\makeatother
\providecommand*\mcitethebibliography{\thebibliography}
\csname @ifundefined\endcsname{endmcitethebibliography}
  {\let\endmcitethebibliography\endthebibliography}{}

\end{document}